\documentclass[sn-nature, pdflatex]{sn-jnl}

\usepackage{graphicx}%
\usepackage{multirow}%
\usepackage{amsmath,amssymb,amsfonts}%
\usepackage{amsthm}%
\usepackage{mathrsfs}%
\usepackage[title]{appendix}%
\usepackage{xcolor}%
\usepackage{textcomp}%
\usepackage{manyfoot}%
\usepackage{booktabs}%
\usepackage{algorithm}%
\usepackage{algorithmicx}%
\usepackage{algpseudocode}%
\usepackage{listings}%
\usepackage{caption}
\usepackage{subcaption}
\theoremstyle{thmstyleone}%
\theoremstyle{thmstyletwo}%

\theoremstyle{thmstylethree}%

\usepackage{setspace}
\begin{document}
\title[]{Observation of a Displaced Squeezed State in High-Harmonic Generation}

\author*[1]{\fnm{David} \sur{Theidel}}\email{david.theidel@polytechnique.edu}
\author[1]{\fnm{Viviane} \sur{Cotte}}
\author[2]{\fnm{Philip} \sur{Heinzel}}
\author[1]{\fnm{Houssna} \sur{Griguer}}
\author[1]{\fnm{Mateusz} \sur{Weis}}
\author[2, 3]{\fnm{René} \sur{Sondenheimer}}
\author[1]{\fnm{Hamed} \sur{Merdji}}

\affil[1]{\orgdiv{Laboratoire d’Optique Appliquée}, \orgname{ENSTA ParisTech, CNRS, École polytechnique}, \orgaddress{\street{828 Boulevard des Maréchaux}, \city{Palaiseau}, \postcode{91120}, \country{France}}}

\affil[2]{\orgdiv{Friedrich-Schiller-University Jena}, \orgname{ Institute of Condensed Matter
Theory and Optics}, \orgaddress{\street{Max-Wien-Platz 1}, \city{Jena}, \postcode{07743}, \country{Germany}}}

\affil[3]{\orgdiv{Fraunhofer Institute for Applied Optics and Precision Engineering IOF}, \orgaddress{\street{Albert-Einstein-Str. 7}, \city{Jena}, \postcode{07745}, \country{Germany}}}

\abstract{\textbf{High harmonic generation is a resource of extremely broad frequency combs of ultrashort light pulses. The non-classical nature of this new quantum source has been recently evidenced in semiconductors by showing that high harmonic generation  generates multimode squeezed states of light. Applications in quantum information science require the knowledge of the mode structure of the created states, defining how the quantum properties distribute over the spectral modes. To achieve that, an effective Schmidt decomposition of the reduced photonic state is performed on a tripartite harmonic set by simultaneously measuring the second- and third-order intensity correlation function. 
The Schmidt number is estimated which indicates an almost single-mode structure for the each harmonic, a useful resource in quantum technology.
By modelling our data with a displaced squeezed state, we retrieve the dependencies of the measured correlation as a function of the high harmonic driving laser intensity. 
The effective high-harmonic mode distribution is retrieved, and the strength of the contributing squeezing modes is estimated. Additionally, we demonstrate a significant violation of a Cauchy-Schwarz-type inequality for three biseparable partitions by multiple standard deviations. Our results confirm  non-classicality of the high-harmonic generation process in semiconductors . The source operates at room temperature with compact lasers, and it could become a useful resource for future applications in quantum technologies.}}


\maketitle

\section*{Introduction}
Non-classical sources of light are a versatile resource in quantum information science and technology \cite{o2009photonic}, thanks to their unique properties like entanglement and squeezing to boost metrology \cite{polino_photonic_2020, schnabel2017squeezed, gessner2020multiparameter}, imaging \cite{defienne_advances_2024, moreau2019imaging}, or establish secure communication \cite{braunstein_quantum_2005-1, gottesman2003secure, yonezawa2004demonstration}. Another promising prospect is photonic quantum simulation \cite{zhong_quantum_2020, cai2017multimode, walmsley_light_2023} that use displaced squeezed states of light, superposed to simulate arbitrary quantum systems. The main advantages of photonic quantum states are their robustness against decoherence due to their low interaction with the environment. Furthermore, the agility of light manipulation offers the possibility of encoding high-dimensional entanglement in time bins, frequency bins, or topological states \cite{erhard2020advances}.

It is important to determine the mode structures of engineered quantum states. The concept of modes is well known in classical electromagnetism and refers to multiple solutions of the Maxwell equations. The electromagnetic field can propagate in e.g., different spatial, polarization, or spectral modes. Here, we are instead investigating the distribution of quantum mechanical properties, which are not necessarily restricted to one optical mode \cite{fabre2020modes}. The mode analysis of quantum states is of great importance, as it leverages engineering to monitor and select favourable modes for specific photonic quantum state applications \cite{ wang2018multidimensional, wright2022nonlinear, caldwell2022time}.

To this extent, higher-order intensity correlation functions can be analyzed to infer the underlying modal distribution of the generated state. 
Being closely connected to the moment-generating functions, higher-order photon number correlation functions probe properties like the width or skewness of the underlying photon number distribution.
In the case of multimode squeezed states of light, they can also, under some assumptions, be used to estimate the modal squeezer distribution \cite{laiho_measuring_2022}. 
Based on this, entanglement can be quantified by an analysis of the multimodal structure through an effective Schmidt mode decomposition. 

In this work, we analyze the modal distribution of a non-classical state of light generated by a new source of quantum light, high harmonic generation (HHG), by measuring higher-order correlation functions. The non-classicality of a tripartite harmonic state is demonstrated through the violation of a multimode Cauchy-Schwarz inequality (CSI). Here, the discrete nature of the HHG light and the strong correlation between photons of different harmonic orders are responsible for the violation of the CSI. Further, we extend existing techniques analyzing a squeezed vacuum state \cite{christ_probing_2011, laiho_measuring_2022, wakui_ultrabroadband_2014}, to a more general one including displacement of the state. 

Theoretical works on HHG show the potential of the process to be employed as a highly versatile source of non-classical light. HHG is a non-perturbative frequency upconversion process that can generate broadband frequency combs covering a spectral range from the near-infrared down to the extreme ultraviolet \cite{krausz2009attosecond, ghimire_strong-field_2014, yue_introduction_2022}. The attosecond time structure of the HHG pulse is inherited from its underlying highly non-linear interaction of the electrons with the periodically oscillating driving electric field potential of an ultrashort pulse of light. This petahertz-frequency electron wave-packet pendulum generates then attosecond bursts of light as a response to the strong field interaction.

Non-classical effects in HHG open new avenues to be explored, exploiting the broadband generation of non-classical photon correlation, multi-partite entanglement \cite{gorlach2020quantum, sloan_entangling_2023} and attosecond control of squeezed states of light \cite{stammer2024entanglement}, as well as conditioned generation of cat states \cite{lewenstein_attosecond_2022,rivera-dean_nonclassical_2024}. The first experimental work on semiconductor HHG (SHHG) reported the generation of squeezed states of light \cite{theidel_evidence_2024} and addressing HHG statistical and spectral properties by driving the process with non-classical light \cite{rasputnyi2024high}.

In this letter, we explore the mode decomposition of SHHG from a cadmium telluride (CdTe) crystal, which has reported a high conversion efficiency \cite{long2023high}, making it favorable for signal-demanding quantum information applications. By simultaneously measuring the second-order correlation function (SCF) and third-order correlation function (TCF), we extract information about the modal structure and parameters of the state generated. Further, we confirm previous results showing multimode squeezed light \cite{theidel_evidence_2024}. The violation of a multimode Cauchy-Schwarz-type inequality (CSI) is shown to be even more significant in CdTe compared to previous studies \cite{theidel_evidence_2024}. 

A displaced squeezed state of light is verified by our measurement supported by numerical simulation. Our analysis shows that, in addition to squeezing, displacement is necessary to explain the observed dependencies of correlation functions on the driving laser intensity. Based on the Schmidt decomposition, the dimensionality of the state is estimated.

\section*{Measurement of the multimode Cauchy-Schwarz inequality}
The experimental demonstration is performed using a fiber-based laser system generating short-wavelength infrared (SWIR), ultrashort femtosecond pulses to excite SHHG in CdTe with [110] crystal cut and select three harmonics from the third to the fifth order (Fig. \ref{fig:cdte_spectra}). The non-perturbative regime of SHHG is monitored by observation of the intensity crossovers, which are clearly visible for all of the harmonic orders (Fig. \ref{fig:intensity_scaling}). Using spectrally selective dichroic mirrors and spectral bandpass filters, we direct each high-harmonic order towards a beamsplitter behind which two single-photon resolving avalanche photodiodes are placed. The full experimental setup is shown in Fig. \ref{fig:g2_setup}. This Hanbury-Brown-Twiss (HBT) like configuration allows us to retrieve
the SCF expressed as

\begin{equation}
g^{(2)}_{ij}(\tau) = \frac{\langle a_i^\dagger(\tau) a_j^\dagger(0) a_i(0) a_j(\tau) \rangle}{\langle a_i^\dagger(0) a_i(0) \rangle \langle a_j^\dagger(\tau) a_j(\tau) \rangle}\qquad ,
\label{eq:g2_broad_function}
\end{equation}
where $a^\dagger$ and $a$ denote the creation and annihilation operators of photons in the mode $i$ or $j$ representing two of the three central spectral modes of three harmonic states H3, H4, and H5.
As the HHG pulses have a duration much shorter than the timing resolution of the detection, we effectively perform a broadband, multimode measurement 

\begin{equation}
g^{(2)}_{ij}(0) \equiv g^{(2)}_{ij} = \frac{\int d \omega_1 \int d \omega_2 \langle \hat{a}_i^\dagger(\omega_1)\hat{a}_j^\dagger(\omega_2)\hat{a}_i(\omega_1)\hat{a}_j(\omega_2)\rangle}{\int d \omega_1 \langle \hat{a}_i^\dagger(\omega_1)\hat{a}_i(\omega_1)\rangle\int d \omega_2 \langle \hat{a}_j^\dagger(\omega_2)\hat{a}_j(\omega_2)\rangle} \,
\label{eq:g2_freq_int}
\end{equation}
where the convolution of the photon statistics of the spectral modes is measured \cite{theidel_evidence_2024, christ_probing_2011, laiho_measuring_2022}. In turn, this acquisition is sensitive to squeezing effects in the radiation. 
We observe a characteristic transition of the SCF when adjusting the driving laser intensity from $g^{(2)}_{ij} \gg 2$ to $g^{(2)}_{ij} = 1$ (Fig. \ref{fig:g2_intensity}). Connecting the SCF to the photon statistics of the radiation, these values indicate a pronounced Super-Poissonian photon number distribution in each of the measured harmonic orders. This distinctive dependence on the driving laser intensity and strong inter-beam correlations are indicative of the presence of two-mode squeezing in the HHG as reported in previous works \cite{theidel_evidence_2024, christ_probing_2011, cardoso2021superposition}. 

\begin{figure}
  \centering
  \includegraphics[width=\textwidth]{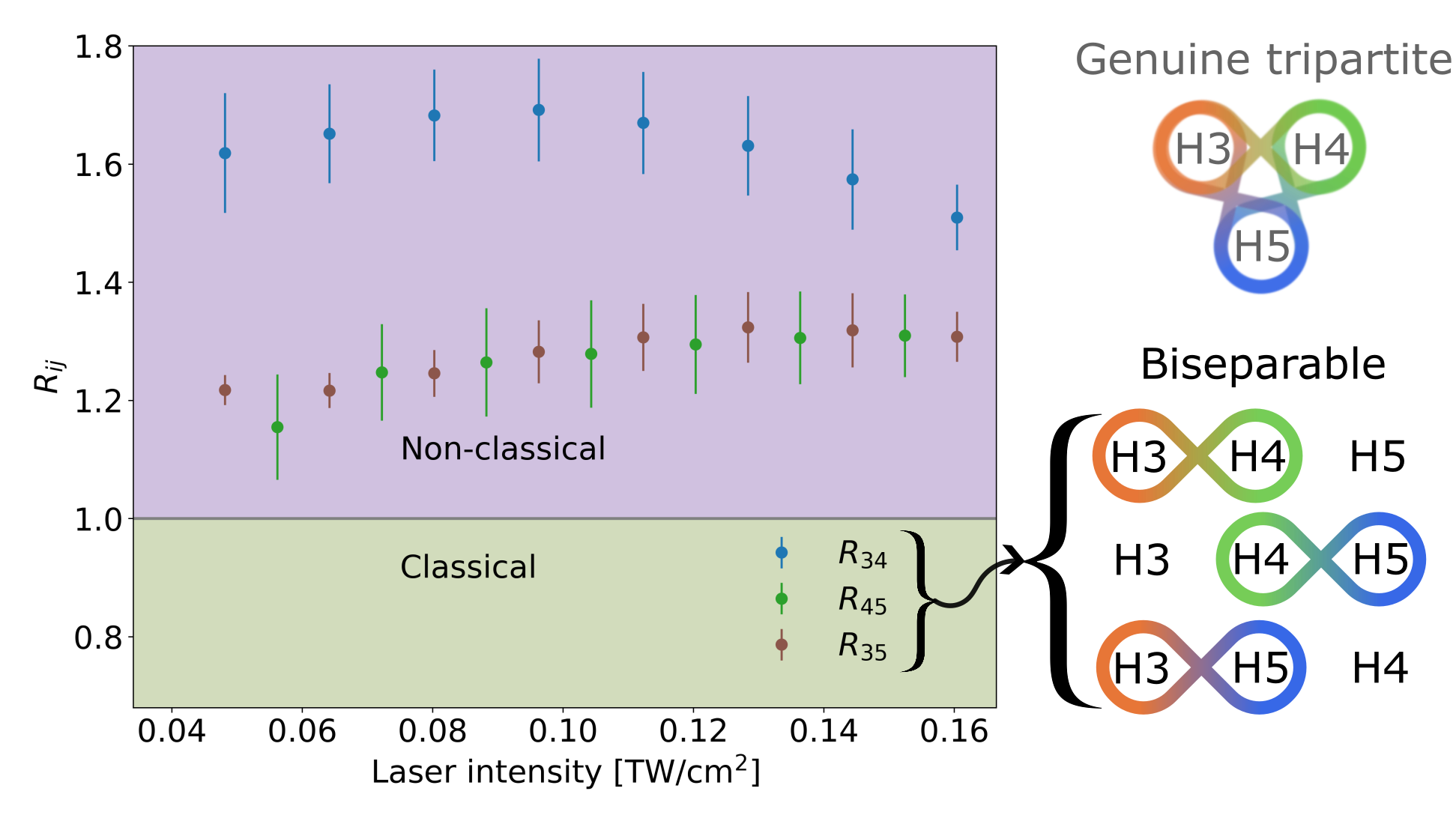}
\caption{From the experimental data in Fig. \ref{fig:g2_intensity} we evaluate the $R$ parameter, given as Eq. \eqref{eq:R_param_exp}. For all bipartite sets at all driving laser intensities we observe $R > 1$ with high significance, confirming that the generated state from SHHG is non-classical. The error bars are calculated from the multiple repetitions as indicated in the experimental data. The pairwise CSI violation and the observation of two-mode squeezing signatures imply the generation of three entangled biseparable states.}
\label{fig:R_param_exp}
\end{figure}

We evaluate a multimode Cauchy-Schwarz inequality expressed as
\begin{equation}
g^{(2)}_{ii} g^{(2)}_{jj} \ge \left[g^{(2)}_{ij}\right]^2
\label{eq:CSI}
\end{equation}
to confirm the presence of non-classical correlations between the harmonic states. We calculate the $R$ parameter defined as 
\begin{equation}
R_{ij} := \frac{\left[g^{(2)}_{ij}\right]^2}{g^{(2)}_{ii} g^{(2)}_{jj}}  
\label{eq:R_param_exp}
\end{equation} 
from the experimental data, such that a violation of the CSI corresponds to $R > 1$. The results are shown in Fig. \ref{fig:R_param_exp}, where we obtain a significant violation for all bipartite sets from the experimental data (complete data set displayed in Fig. \ref{fig:g2_intensity}). The CSI is known to be violated from two-mode squeezed states, as shown by theory \cite{peng_introduction_1998} and experiment and proves the presence of non-classical correlations between the analysed modes \cite{wasak_cauchyschwarz_2016}. Indeed, the bipartite reductions correspond most likely to a two-mode squeezed state, exhibiting entanglement between modes.
 The tripartite state, as illustrated in Fig. \ref{fig:R_param_exp}, can therefore be interpreted as three entangled biseparable states.
Genuine tripartite entanglement is out reach here it requires entanglement measures beyond the current study \cite{cobucci2024detecting}.

\section*{Estimation of the mode distribution}
Further, insights in the modal structure of the generated squeezed state of light can be obtained by simultaneous measurement of the SCF and TCF \cite{christ_probing_2011, laiho_measuring_2022}. The broadband TCF is given as

\begin{equation}
g^{(3)}_{ijk} = \frac{\int d \omega_1 \int d \omega_2 \int d \omega_3 \langle \hat{a}_i^\dagger(\omega_1)\hat{a}_j^\dagger(\omega_2)\hat{a}_k^\dagger(\omega_3)\hat{a}_i(\omega_1)\hat{a}_j(\omega_2)\hat{a}_k(\omega_3)\rangle}{\int d \omega_1 \langle \hat{a}_i^\dagger(\omega_1)\hat{a}_i(\omega_1)\rangle\int d \omega_2 \langle \hat{a}_j^\dagger(\omega_2)\hat{a}_j(\omega_2)\rangle \int d \omega_3 \langle \hat{a}_k^\dagger(\omega_3)\hat{a}_k(\omega_3)\rangle} .
\label{eq:g3_broad_function}
\end{equation}

Being closely linked to the generating moments, measuring higher order correlations can give further insights into the underlying photon number probability distribution function.
Indeed for squeezed vacuum states the analysis of the simultaneous measurement of $g^{(2)}$ and $g^{(3)}$ can be used to estimate the broadband squeezing distribution, the squeezing strength and the effective cooperativity parameter K \cite{christ_probing_2011, wakui_ultrabroadband_2014}. 

\begin{figure}
\centering
\includegraphics[width=1\textwidth]{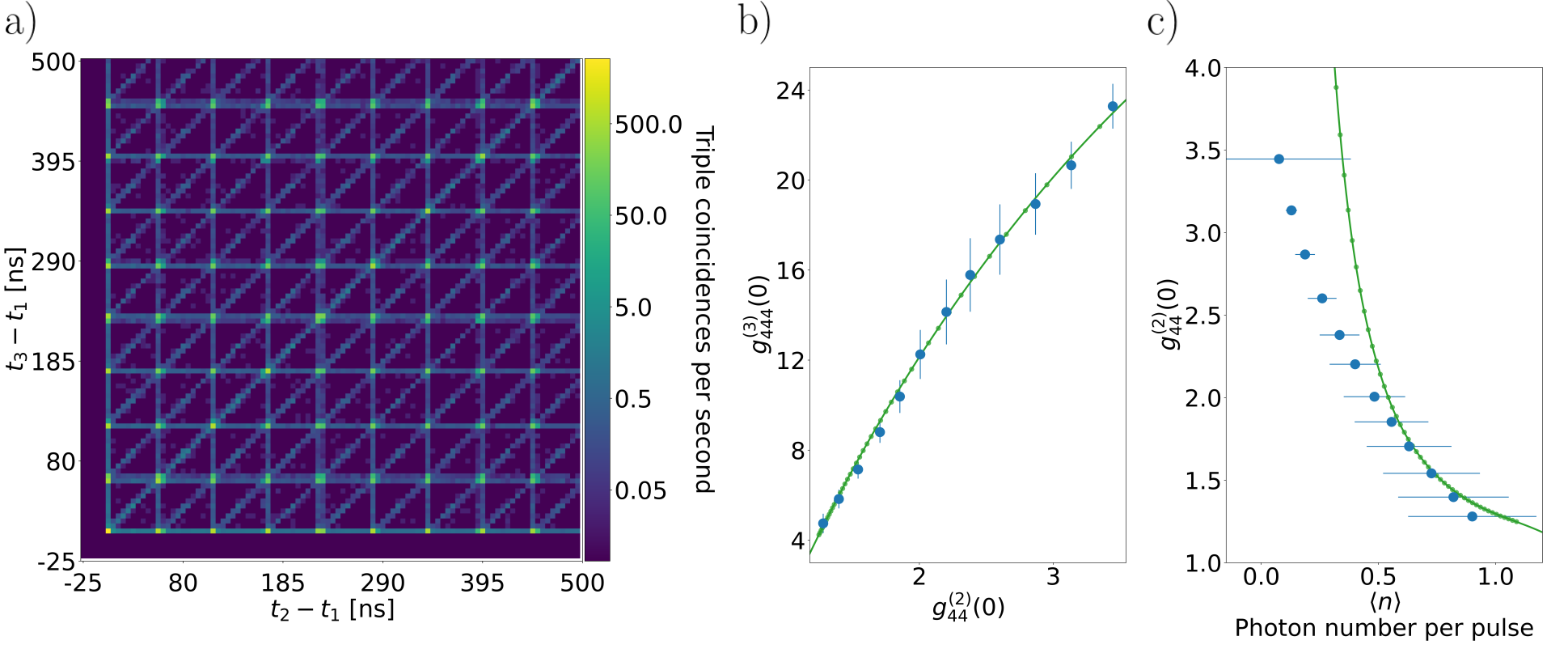}

\caption{Figure a) shows the raw dataset of the triple coincidence counts for the fourth harmonic using the setup in \ref{fig:g3_setup}. The colormap encodes the relative number of triple coincidences per second. The periodic grid appears due to the pulsed excitation. The spacing between the peaks coincides with the time difference between the ultrashort laser pulses used for the excitation of the SHHG process. We calculate the TCF by normalizing the raw three-fold coincidences with respect to the mean height of the peaks. Peaks which are placed on the diagonal and as well as the peaks in the first row and first column are not taken into account for then normalization, as they may contain correlations from photons generated in the same pulse, and thus might not be uncorrelated. The figures b) and c) show the results from the joint measurement of the SCF and TCF in a single-beam setup (Fig. \ref{fig:g3_setup}) for the fourth harmonic depicted in blue. An almost linear relationship between $g^{(2)}_{44}$ and $g^{(3)}_{444}$ is apparent in Figure a). Figure b) shows the SCF $g^{(2)}_{44}$ over the loss-corrected mean photon number per pulse $\langle n \rangle$. The green curve is the result from a simulation of a multimode, displaced squeezed thermal state from which the respective observable are calculated. The simulation reproduces well the measured values and suggest that the SHHG state is a displaced squeezed state. A similar analysis is shown for the dataset of the fifth harmonic in Fig. \ref{fig:g2_g3_H5}.}
\label{fig:g2_g3_H4}
\end{figure}

The adapted experimental setup is shown in Fig. \ref{fig:g3_setup}. We select a single harmonic order to a single-beam configuration and add an additional beamsplitter behind the dichroic mirror. At the three outputs of this configuration, the single-photon avalanche diodes are placed and connected to the time tagging device to acquire the two-fold and three-fold coincidence counts to simultaneously acquire the SCF and TCF.

By normalizing the raw triple coincidences (Fig. \ref{fig:g2_g3_H4}a)) to the height of the off-diagonal peaks, we calculate the third-order correlation function $g^{(3)}_{iii}$. Details on the data treatment can be found in the methods section. By changing the driving laser intensity, we obtain a set of data with distinctive values for intrabeam correlations $g^{(2)}_{ii}$ and $g^{(3)}_{iii}$.  
Similar to the analysis carried out in \cite{christ_probing_2011, wakui_ultrabroadband_2014}, we establish a linear relationship between these two parameters (Fig. \ref{fig:g2_g3_H4} and Fig. \ref{fig:g2_g3_H5}), from which we estimate the squeezing strength and distribution.
As a result, the generated state can be interpreted as a displaced squeezed state in contrast to the squeezed vacuum state analyzed in \cite{christ_probing_2011, wakui_ultrabroadband_2014}. 

\begin{figure}
\centering
\includegraphics[width=\textwidth]{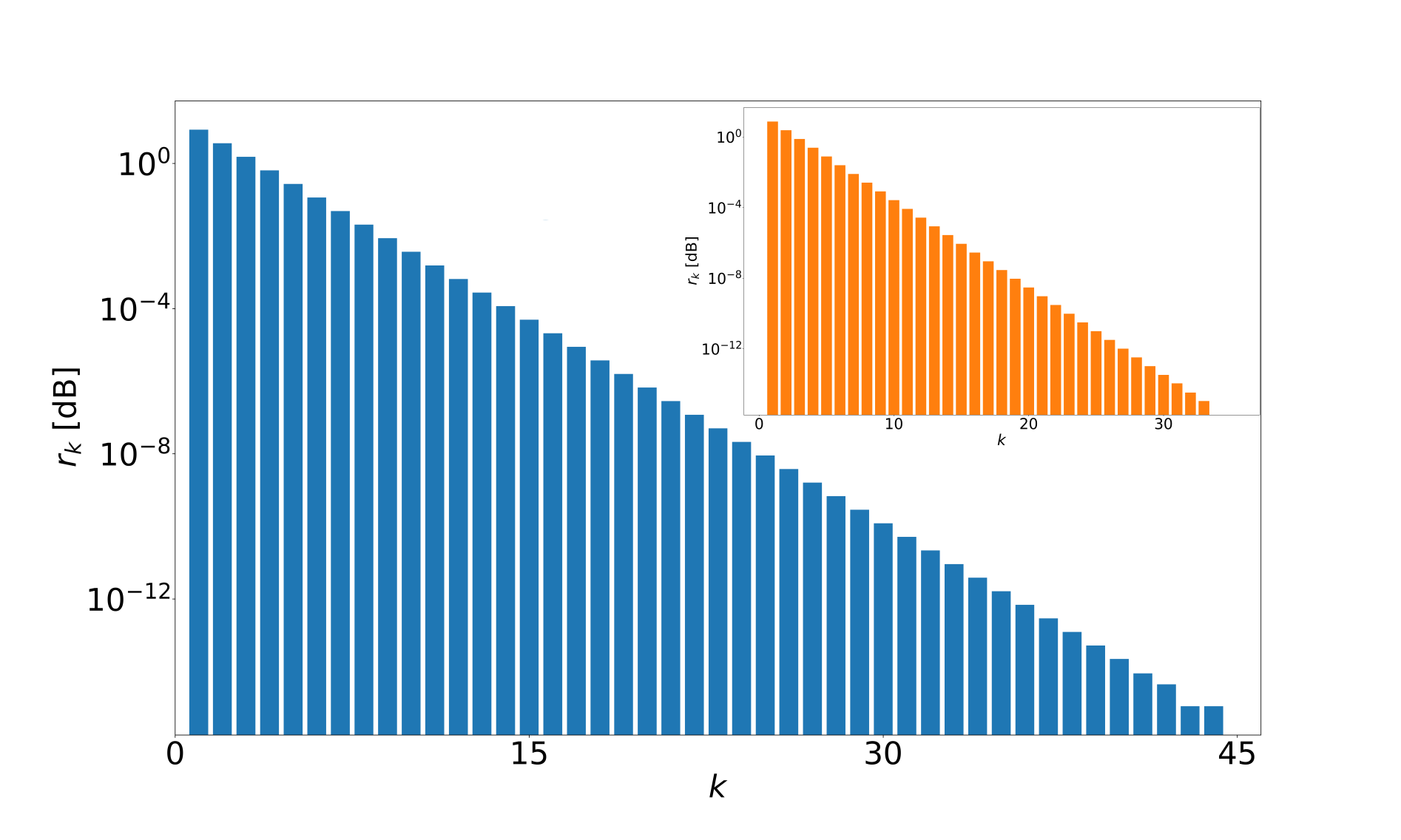}
\caption{Estimated squeezer distribution for the fourth (blue) and fifth harmonic (orange), calculated from a datapoint of the simulated multimode, displaced squeezed states from Fig. \ref{fig:g2_g3_H4} and Fig. \ref{fig:g2_g3_H5}. Multiple modes exhibit non-neglible squeezing with a maximum squeezing level of 8.5 dB for the fourth harmonic and 7.53 dB for the fifth harmonic. Additionally, we calculate the effective Schmidt numbers to be $K_{\text{H4, eff}} = 1.43$ ($K_{\text{H5, eff}} = 1.23$). }
\label{fig:sq_distb_H4}
\end{figure}

Motivated by the interaction Hamiltonian modelling the SHHG process \cite{theidel_evidence_2024}, we base
our analysis on a simulated multimode system of a multimodal Gaussian state. In the most general case, this includes multimode displacement, squeezing and rotation of the vacuum in case of a pure state. However, for the purpose of modelling the state, we neglect rotations in a first analysis, since they can be switched past squeezing and displacement operators to act on the vacuum first \cite{ma1990multimode}. The actual state generated might be more refined with rotations present among the modes. However, this aspect is beyond the scope of our current explorative analysis. We would like to emphasize at this point that more detailed theory investigations, beyond the current state of the art of such systems, are necessary to properly determine the impact of such effects.
Consequently, we use the fact that the vacuum is an eigenstate of a rotation with eigenvalue 1 which leaves us with
only a squeezing and displacement operation. For simplicity, we have also forgone the inclusion of two beam squeezing between spectral modes for now, such that the squeezing matrix is diagonal, and the operator can be decomposed into $k$ single beam squeezer. In order to take losses into account, we do not restrict the subsequent analysis to pure states (squeezing and displacement) but allow for a general mixed Gaussian state, i.e., a displaced, squeezed thermal state \cite{yuan2015squeezed}.

The density matrix of such a state goes along the following standard formulation:
\begin{equation}
 \rho_k = \bigotimes_{k=1}^d \hat{D}_k(\alpha)\hat{S}_k(\zeta) \rho_{\text{th}, k}\hat{S}_k^\dagger(\zeta)\hat{D}_k^\dagger(\alpha)
 \label{eq:rho}
\end{equation}
where $\hat{D}_k(\alpha)=\exp{(\alpha_k \hat{a_k}^\dagger - \alpha_k^* \hat{a_k}})$ 
is the displacement operator,
$\hat{S}_k(\zeta)=\exp{ \lbrace (\zeta_k^* (\hat{a_k}^\dagger)^2-\zeta_k \hat{a_k}^2)/2 \rbrace }$ 
the squeezing operator with $\zeta_k=r_k\cdot\exp{(i \theta_k)}$ and $\rho_{\mathrm{th}, k}=\sum_n P(n_k)|n\rangle\langle n|$ the density matrix of a thermal state with occupation number $n_k$ and $d$ the number of spectral modes constituting the corresponding SHHG mode.
We discretize the integrals in equation Eq.\,\eqref{eq:g2_broad_function}
\begin{equation}
g^{(2)} \approx \frac{\sum_l^d \sum_m^d \langle \hat{a}_l^\dagger \hat{a}_m^\dagger \hat{a}_l\hat{a}_m\rangle }{\sum_l^d \langle \hat{a}_l^\dagger\hat{a}_l \rangle \sum_m^d \langle \hat{a}_m^\dagger \hat{a}_m \rangle}
\end{equation}
and assume uncorrelated squeezer modes. This simplifies the numerator in the case {$l \neq m$} to products of expectation values of photon number operators. The integral expression Eq. \eqref{eq:g3_broad_function} is discretized analogously. Further, we parametrize the squeezing parameter as $r_k = B\cdot \lambda_k$ with $\lambda_k = \sqrt{1 - \mu^2}\mu^k$, assuming a 2D Gaussian joint-spectral density distribution (JSD) \cite{christ_probing_2011} with an underlying, normalized thermal squeezer distribution characterized by the thermal occupation parameter $\mu$. The assumption of a Gaussian JSD is justified through the use of narrowband spectral filters \cite{zielnicki2018joint}.

The strength of the process $B$ is assumed to be isotropic for all modes $k$ as well as isotropic displacement over the modes $\alpha_k = \alpha$ and $n_k = n$. Finally $\theta_k = 0$. 

We note that these assumptions are primarily made to simplify the simulations in accordance with \cite{christ_probing_2011, wakui_ultrabroadband_2014}. However, they may still serve as a rough estimation of the underlying parameters, as the state generated in SHHG is likely Gaussian in a first approximation.

In total, the parameters $B$, $\mu$, $\alpha$ and $n_{\mathrm{th}}$ are employed to create $k$ states specified by Eq.\,\eqref{eq:rho}. 

 Different values for the SCF, TCF and mean photon number $\langle n \rangle$ are obtained by subsequently including up to $d=50$ modes to the simulation \cite{johansson2012qutip}. The convergence of the simulation is verified by decreasing the number of modes by 15 and confirming that the same parameters and estimated Schmidt number is obtained. Additionally the simulated data points, shown as green points in Fig. \ref{fig:g2_g3_H4}b) and \ref{fig:g2_g3_H4}c), show excellent convergence to the experimental data with an increase of k.
The parameters for both datasets on the fourth and fifth harmonics are found by minimizing the root mean square deviation between the data and the obtained values from the simulation on the $g^{(2)}$ vs. $g^{(3)}$ and $\langle n \rangle$ vs. $g^{(2)}$ dataset. The results for the fourth and fifth harmonic are shown in Fig. \ref{fig:g2_g3_H4} and Fig. \ref{fig:g2_g3_H5}. The simulation reproduces the experimentally observed dependence well. For both measured harmonic orders we find, that a state with non-negligible squeezing is needed to interpret the experimental data. In the case of the fourth harmonic, the optimal parameters are found to be $B_{\text{opt}, 4} = 0.471$, $\mu_{\text{opt}, 4} = 0.4226$, $\alpha_{\text{opt}, 4} = 0.127$ and $n_{\text{th},\text{opt}, 4} = 0.001$. For the fifth harmonic $B_{\text{opt}, 5} = 0.397$, $\mu_{\text{opt}, 5} = 0.32$, $\alpha_{\text{opt}, 5} = 0.161$ and $n_{\text{th},\text{opt}, 5} = 0.001$ . From the numerical analysis we estimate the effective Schmidt number $K_{\text{eff}} = 1 / \sum_k \lambda_k^4$ \cite{eberly2006schmidt} as $K_{\text{H4, eff}} = 1.43$ and $K_{\text{H5, eff}} = 1.23$.

Despite restrictive parameter assumptions, we find the simulations of the model to be in good agreement with the experimental data and fitting well the interpretation of the observed scaling of the correlation functions with the driving laser intensity due to the broad squeezer distribution, as previously suggested in \cite{theidel_evidence_2024}.
The deviations of the experimental data, could originate from the finite data size introducing a statistical uncertainty, which becomes more pronounced for a low number of mean photons \cite{wakui_ultrabroadband_2014}. Still, our interpretation and numerical analysis provides a good explanation. As the SHHG process is highly complex and the theoretical work on understanding the emergence of quantum features has just recently started, we expect that other intrinsic effects such as light-matter entanglement might contribute to the observed features of the data \cite{rivera2022light}.

\section*{Discussion}
In summary, we present a thorough analysis of the second and third-order photon number correlation function of the state created in SHHG. In a three-beam configuration, we measure the violation of the CSI in the bipartite partitions, a strong signature for non-classical correlations. This result motivates for an investigation of genuine tripartite entanglement in SHHG.  Indeed, in quantum information science, especially computing, a central goal is to generate fully controllable entangled states featuring increasingly many ($N>2$) particles. SHHG could potentially meet this key requirement for quantum advantage.
We restrict here our quantum correlations studies to three biseparable states and further investigate for a specific harmonic the behaviour of the second- and third-order correlation function with the driving laser intensity.
The data are representative of a displaced squeezed thermal state. On the basis of the numerical analysis, a Schmidt decomposition is performed to quantify the mode structure of the state generated and estimate the effective Schmidt number and squeezing strength of the spectral modes.
This result confirms and extends previous work reporting two-mode squeezed states from SHHG \cite{theidel_evidence_2024}.
The presence of displacement and squeezing of the SHHG state, show the complexity and future prospects for fine control of the state generated using advanced quantum engineering. Spectrally entangled states are major resources in quantum information science and emerging technologies such as quantum metrology, computing, and cryptography.

 Our findings further establish the application of quantum optical methods to the tripartite SHHG state and are, to the best of our knowledge, a unique study of the squeezer distribution generated from SHHG. 
 With this, we pave the way for further studies and applications of the displaced squeezed state generated via SHHG.
 This further establishes SHHG as a highly promising candidate for applications in the broad field of quantum state engineering, gaining possible quantum advantage due to the potential high dimensionality of the HHG bosonic platform \cite{yi2024generation}. 

\section*{Methods}
Our experimental setups are based on a 200 mW all-fiber pulsed laser system with a central wavelength of 2100 nm and a repetition rate of 18.66 MHz. The ultrashort laser pulses with a duration of 80 fs are linearly polarized and focused on the CdTe[110] crystal with a focal length of 2.5 cm. Before, a half-wave plate and polarizer combination allowed us to adjust the driving laser intensity. After the sample, a slightly closed aperture and infrared filter separate the driving laser light from the generated high-harmonic orders. An additional broadband polarizer removes the luminescence background of the SHHG signal. The SHHG radiation is collimated and the harmonic modes 3 to 5 spatially separated by means of dichroic mirrors and spectral filters. For the measurement of the SCF, a beamsplitter is placed behind each spectral filter, and the two outputs are focused on single photon avalanche diodes (SPADs). They are connected to a time-to-digital converter (TDC), recording the arrival time of the photons registered on the SPADs. From the time-stamps, the mean photon number as well as intra- and interbeam correlation are calculated. To record the threefold coincidence counts, a single harmonic order is selected, and two beamsplitters are placed behind the spectral filter. Behind the three beamsplitter outputs, the SPADs are placed and threefold coincidence recorded via the TDC. At the same time, the two outputs of the second beamsplitter are used to record coincidence counts. In both setups, care is taken to not saturate the binary detectors by having an event rate of less than one photon per pulse.
\subsection*{Data treatment and simulation}
Assuming that coincidence events from photons created in subsequent pump pulses are uncorrelated, yielding $g^{(2)} = 1$, we normalize the raw coincidence counts with respect to the satellite peaks. For the calculation of the SCF, Gaussian peaks with a coefficient of determination of at least 90 $\%$. In total, the normalization is then performed over 40 satellite peaks with a percentage coefficient of variation of the root mean square deviation (RMSD) of approximately 15 $\%$. The maximum value of a fitted Gaussian to the central peak is used as the final value of $g^{(2)}(0)$. The error bars are calculated from around 10 repetitions of the experiment. 

A similar approach is followed to obtain the TCF. The raw three-fold coincidence counts shown in Fig. \ref{fig:g2_g3_H4}a) are normalized with respect to the satellite peaks, as $g^{(3)} = 1$ for uncorrelated events. Only off-diagonal satellite peaks are accounted for, and the first row and first column are neglected as well. For these, photon events from the same laser pulse could create a threefold coincidence event. A peak-finding algorithm is used to identify the maxima and, again, about 50 satellite peaks with a total coefficient of variation of less than 15 $\%$. The value $g^{(3)}(0)$ is the read as the maximum value of the central peak in the normalized dataset. The error bars for $g^{(3)} = 1$ are likewise calculated as the standard deviation over the same number of repetitions.

For the numerical simulation, the QuTiP Python framework is employed. A set of $k$ displaced squeezed thermal states is specified with the parameters mentioned in the main text. An optimization algorithm is used to find the parameters that approach our dataset the best. To the right, the mean square error is calculated as the linear best fit to the experimental data, to the linear best fit of the simulated values. For the $g^{(2)}$ vs. $\langle n \rangle$ dataset, a power fit is used instead. The total error to minimize is the sum out of these two errors. In this sense, we optimize the parameters for both datasets simultaneously.

\backmatter

\bmhead{Supplementary information}
The supplementary information contains Figures \ref{fig:cdte_spectra} - \ref{fig:g2_g3_H5}.

\bmhead{Acknowledgements}

 We acknowledge support on the single photon detection by Milutin Kovacev from the Institute of Quantum Optics Hanover. We thank the participants of the Extreme Quantum Optics Workshop 2024 for discussions. P.H. and R.S. acknowledge funding by the Deutsche Forschungsgemeinschaft (DFG, German Research Foundation) – Project-ID 398816777 – SFB 1375 NOA.


\section*{Declarations}


\subsection*{Contributions}
D.T. conceived the study and performed the experiments. D.T., V.C. and P.H. analyzed the data.  D.T., P.H. and R.S. conducted the theoretical analysis. D.T. wrote the simulation. D.T. provided the plots. D.T., H.M., P.H. and R.S. wrote the manuscript with input from all authors. H.M. initiated the quantum HHG research program and acquired the funding. All authors discussed the results. 

\subsection*{Competing interests}
The authors declare no competing financial interests.

\newpage
 
\bibliography{sn-bibliography}

\begin{appendices}

\renewcommand{\thefigure}{S\arabic{figure}}
\renewcommand{\theHfigure}{S\arabic{figure}}
\setcounter{figure}{0}
\newpage

\begin{figure}
  \centering
  \includegraphics[width=0.7\linewidth]{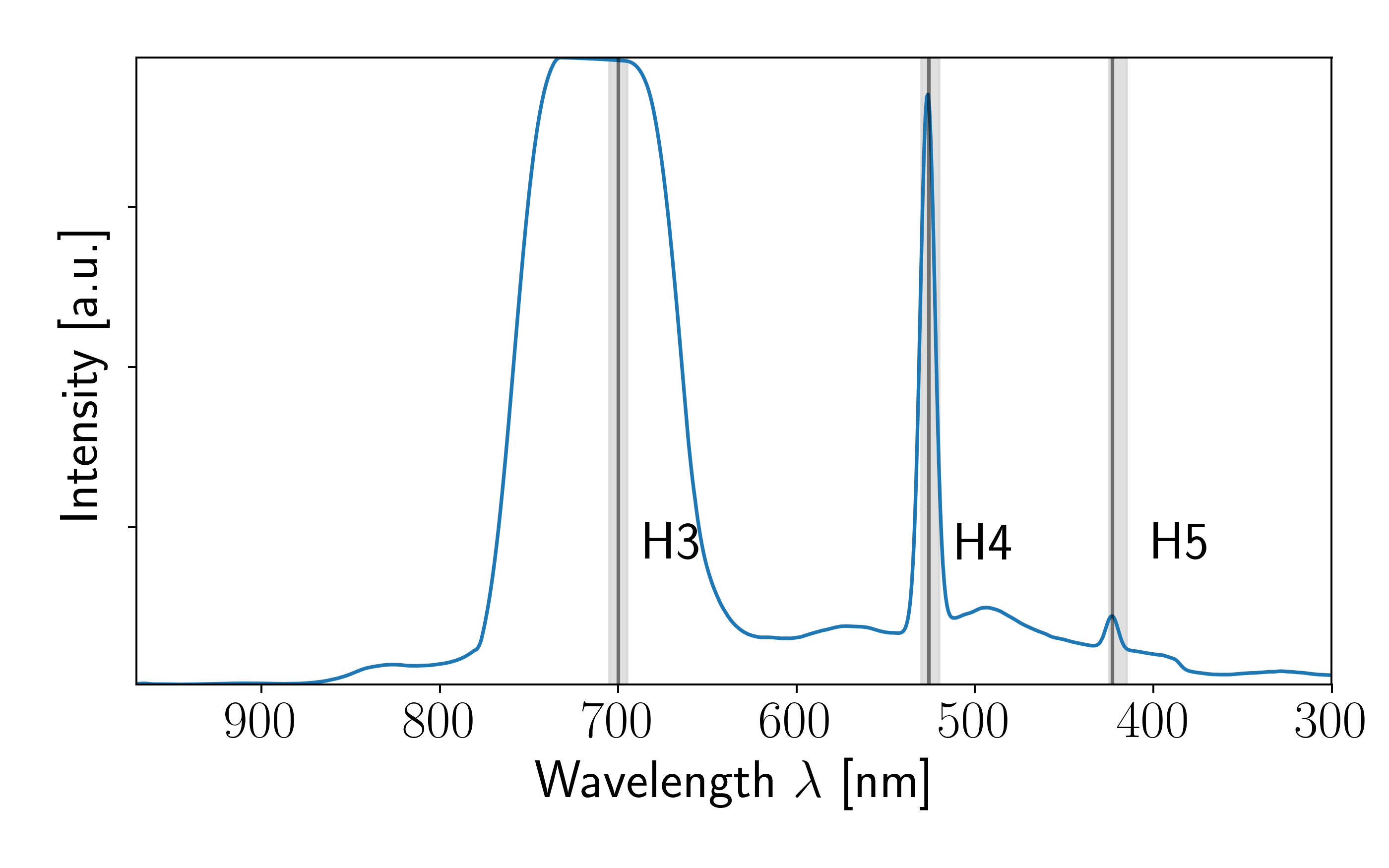}
      \caption{The spectrum shows the three investigated high-harmonics. The grey area represents the width of the spectral filters employed in the experimental setup. Spectral filtering is already a mode selection, as we thereby reduce the width of the squeezer distribution.}
  \label{fig:cdte_spectra}
\end{figure}%

\begin{figure}
     \centering
 \includegraphics[width=\linewidth]{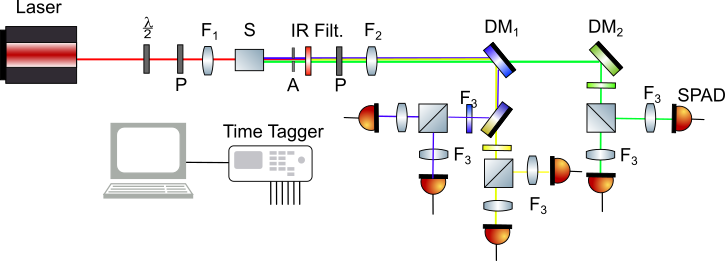}
       \caption{Experimental setup for measuring the SCF $g^{(2)}_{ij}$. The ultrashort pulses are directed through a half-wave plate followed by a polarizer to control the driving laser intensity. The near-infrared pulses are focused on the CdTe[110] sample in which the SHHG process is induced. An aperture and infrared filter are placed behind the sample to filter the pump light. A second polarizer is placed to filter the luminescent background created in the sample. The three harmonic orders are collimated with an achromatic doublet and directed towards the dichroic mirrors to spatially separate the different frequencies. Behind the spectral filters, a beamsplitter is placed and the photons are measured with single photon avalanche diodes. The detectors are connected to a time-to-digital converter to record the arrival times of the photons, from which the SCF is calculated. }
      \label{fig:g2_setup}
\end{figure}

\begin{figure}
  \centering
  \includegraphics[width=\linewidth]{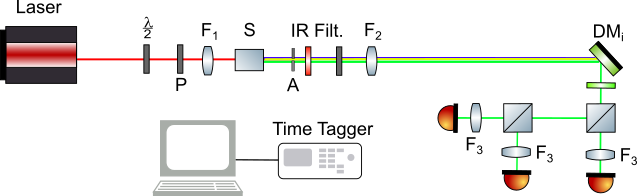}
 \caption{The experimental setup to measure the SCF and TCF in a single beam configuration. The setup is similar to Fig. \ref{fig:g2_setup} up to the collimation lens. Here, only one high-harmonic order is selected with a dichroic mirror and a spectral filter. Behind the spectral filter, two beamsplitters are placed and the radiation at the three outputs of the beamsplitter focused onto the single photon avalanche diodes. These are connected to a time-to-digital converter to record the arrival times of the photon and calculate the SCF as well as TCF.} 
  \label{fig:g3_setup}
\end{figure}

\begin{figure}
  \centering
  \includegraphics[width=0.7\linewidth]{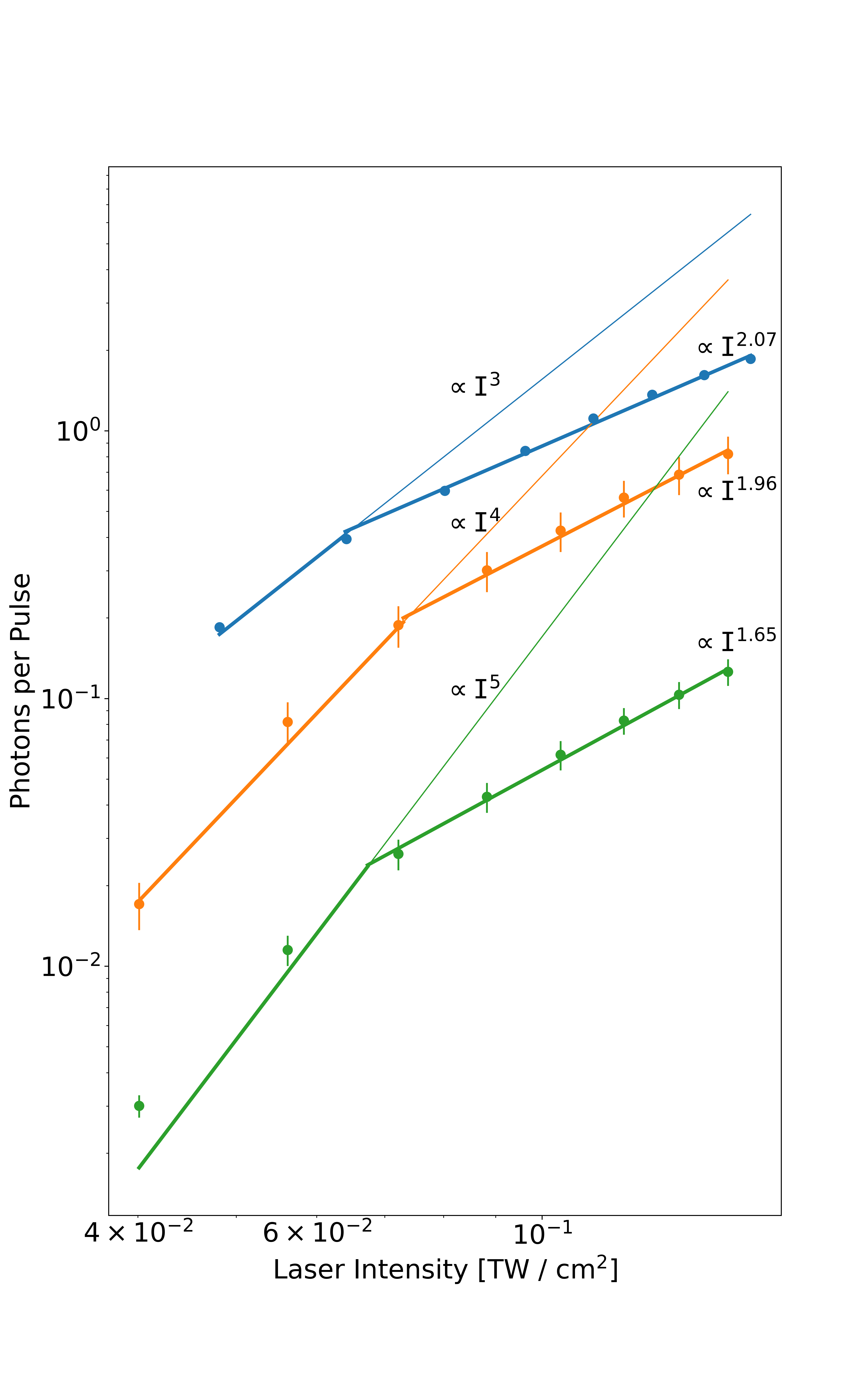}
      \caption{The graph shows the loss-corrected number of photons per pulse over the driving laser intensity. For all recorded high-harmonic orders an intensity crossover is apparent, at which the dependency changes from a pertubative to a non-perturbative scaling law. The fitted curves are a best fit of a power law and the perturbative scaling law is depicted to ease the comparison.}
  \label{fig:intensity_scaling}
\end{figure}%

\begin{figure}
  \centering
  \includegraphics[width=0.9\linewidth]{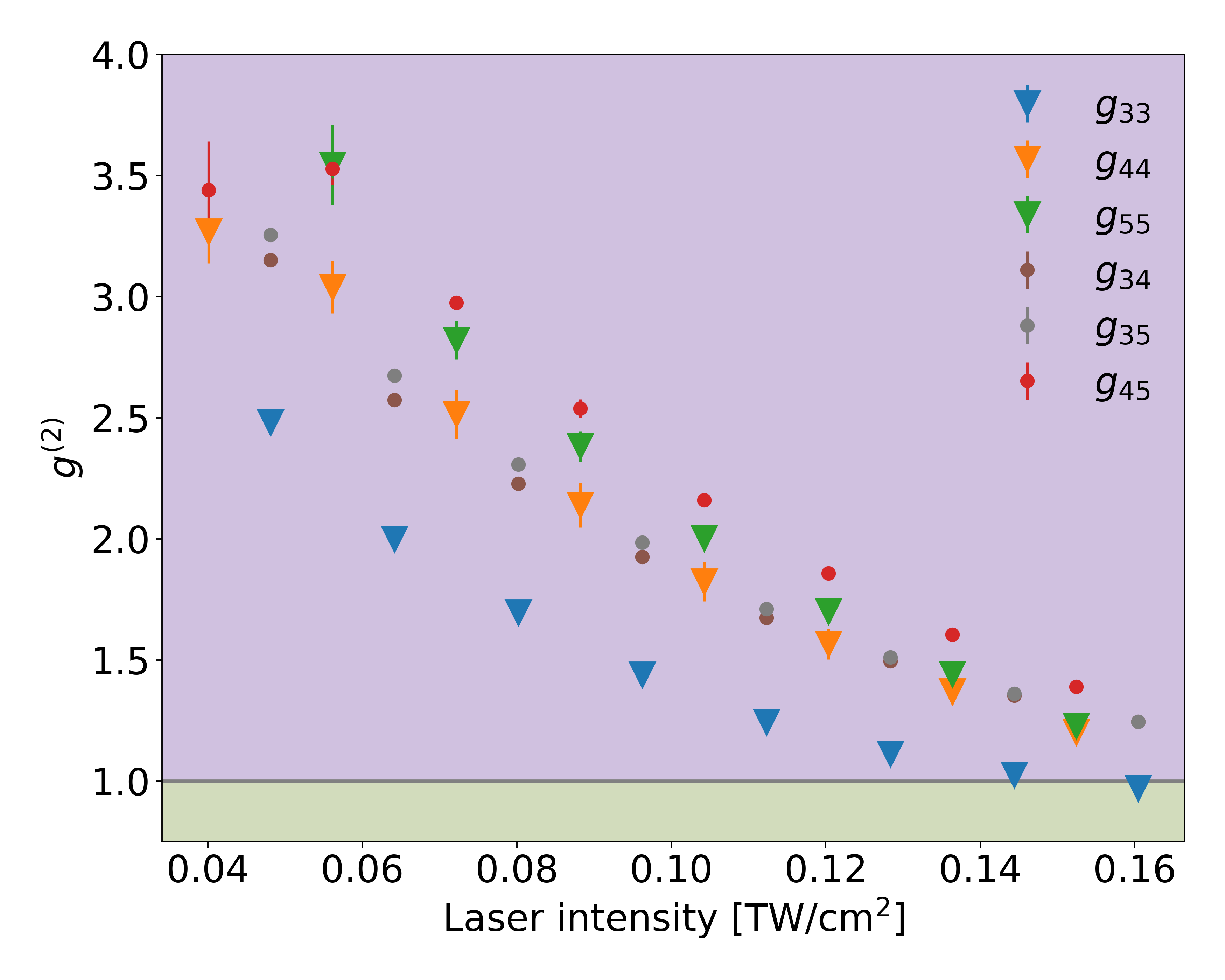}
  \caption{The graphic depicts the auto- and cross-correlations $g^{(2)}_{ij}$ for the three different harmonic acquired with the setup shown in Fig. \ref{fig:g2_setup}. A clear transition from super-bunching with $g^{(2)} > 2$ towards Poissonian photon statistics  $g^{(2)} = 1$ is visible. This behaviour is characteristic for a photonic state exhibiting multimode squeezing. As the driving laser intensity is increased, the squeezer distribution is broadened and a convolution of the photon statistics of the individual spectral modes is measured. As these are uncorrelated with each other, it results in a transitions towards $g^{(2)} = 1$. For low driving laser intensities, only few squeezing modes are excited such that the Super-Poissonian photon number distribution  is resolved.}
  \label{fig:g2_intensity}
\end{figure}

\begin{figure}
\centering
\includegraphics[width=0.9\linewidth]{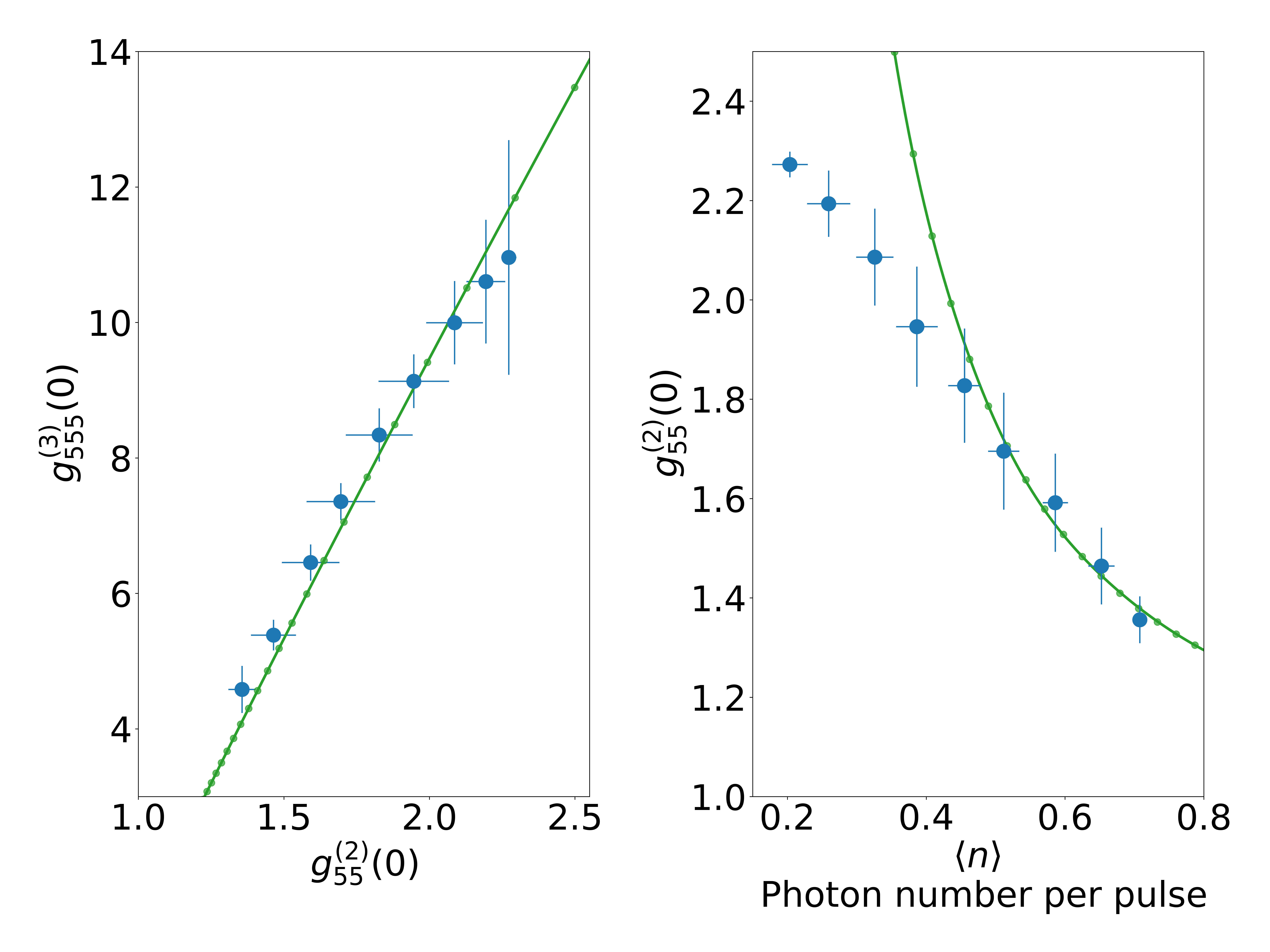}
\caption{Similar to Fig. \ref{fig:g2_g3_H4}, the measured SCF over TCF at time delay zero is shown as the blue points in the left figure. The right graph depicts the SCF over the mean photon number per pulse. Again, an optimization algorithm is employed to retrieve the parameters, for which a displaced squeezed thermal state reproduces the observed behaviour. The result of the simulation is shown as the green curves. The trend is reproduced well by the numerical comparison.}
\label{fig:g2_g3_H5}
\end{figure}




\end{appendices}


\end{document}